\documentclass[12pt]{article}
\usepackage{amsfonts,theorem,amssymb}

\newcommand{\R}{{\mathbb{R}}}

\newcommand{\CP}{{\mathbb{C}}{{P}}}

\newcommand{\beq}{\begin{equation}}
\newcommand{\eeq}{\end{equation}}
\newcommand{\bea}{\begin{eqnarray}}
\newcommand{\eea}{\end{eqnarray}}
\newcommand{\ra}{\rightarrow}

\newcommand{\cd}{\partial}
\newcommand{\wt}{\widetilde}

\newcommand{\rat}{{\sf Rat}}

\newcommand{\ol}{\overline}

\theoremstyle{plain}

{\theorembodyfont{\rmfamily}

}

\begin{document}

\title{Homotopy classification of multiply based textures}
\author{
J.M. Speight\thanks{E-mail: {\tt j.m.speight@leeds.ac.uk}}\\
School of Mathematics, University of Leeds\\
Leeds LS2 9JT, England
}

\date{}
\maketitle

\begin{abstract}
It is shown that the homotopy classification of textures defined on
physical domains with multiple ends at infinity reduces to that of
textures on compact domains if the target space is simply connected. 
The result is applied to the $O(3)$ sigma model on a cylinder $S^1\times
\R$, recently studied by Rom\~ao.
\end{abstract}

\maketitle

Consider a classical field theory defined on space $X$ 
(space-time $X\times\R$) with a field taking values in some target space
$Y$. If $X$ is noncompact with a connected end at infinity, $\cd X$, the
theory may support topological solitons generically called textures if there
exist homotopically nontrivial maps $X\cup\{\infty\}\ra Y$, 
where $X\cup\{\infty\}$ is the one point
compactification of $X$. 
The point is that, either because the vacuum manifold of the
theory is discrete, or because the gradient energy of the field must
converge (or both), one imposes the boundary condition that the field 
$\phi(x)$
approach a single fixed value in $Y$ as $x$ approaches $\cd X$, so one may, 
for topological purposes treat $\cd X$ as a single extra point attached to 
$X$, called $\infty$. The different topological sectors of the theory are
simply the different homotopy classes of maps $X\cup\{\infty\}\ra Y$, where
homotopy means based homotopy, that is, homotopy through maps which
leave $\phi(\infty)$ fixed. 
(For the sake of brevity, map will always mean continuous map.)
It is well known that the based homotopy classes
of maps into $Y$ coincide with the free homotopy classes if $Y$ is 
simply connected. Examples are $O(3)$ sigma model lumps on $\R^2$ and
Faddeev-Hopf solitons on $\R^3$ which are classified by their Brouwer and
Hopf degrees respectively.

There are situations in condensed 
matter physics and cosmology in which $\cd X$ is not connected, but rather
space has several disconnected ends at infinity. A motivating example
is the $O(3)$ sigma model on a cylinder $S^1\times\R$, where $\cd X$ has
two connected components
\cite{rom}. In this case, we may consider physical space
to be a connected compact manifold, which we will continue to denote $X$,
together with an ordered $k$-tuple of points $
(p_1,p_2,\ldots, p_k)$, each representing a {\em component}
 of the boundary at 
infinity. For textures on a cylinder, for
example, $X$ would be $S^2$ with two marked points. 
Let $Y^X$ denote the space of maps $X\ra Y$ given the
compact-open topology. Let $
(q_1,q_2,\ldots,q_k)$ be an ordered $k$-tuple
of  marked
points in $Y$, possibly not all distinct. These points are the boundary
values assigned to the field and may be chosen from a 
discrete vacuum manifold in $Y$, or be arbitrary,
but dynamically frozen. This depends on the dynamical details of the theory,
which are not relevant to the present discussion.
 We shall say a map $f:X\ra Y$ is multiply
 based if $f(p_i)=q_i$ for all $i=1,2,\ldots,k$. This is
the topological condition modelling the finite energy criteria for
textures on space $X\backslash\{p_1,\ldots,p_k\}$.  Denote the
subset of multiply based maps $(Y^X)_*\subset Y^X$. Two maps $f,g\in Y^X$ 
are homotopic, $f\sim g$, if they are path connected in $Y^X$. Similarly,
two maps $f,g\in (Y^X)_*$ are multiply based homotopic, $f\sim_* g$ if they
are path connected in $(Y^X)_*$. Clearly $f\sim_* g$ implies $f\sim g$,
but the converse is, in general, false.
The point is that two multiply based maps can be homotopic through some
homotopy which violates the basing conditions when no homotopy preserving
those conditions exists. The purpose of this letter is to demonstrate that
this cannot happen if $Y$ is simply connected:
\vspace{0.1cm}

\begin{center}
{\it Let $f,g\in (Y^X)_*$ and $f\sim g$. If $\pi_1(Y)=0$, then $f\sim_* g$.}
\end{center}
\vspace{0.1cm}

\noindent
Thus, the only extra
homotopy invariants possessed by textures in such
theories are the combinatorial ones arising from the choice of boundary 
values defining the basing conditions. 
This generalizes the
standard result that (singly) based maps are (singly) based homotopic if and
only if they are free homotopic, when $Y$ is simply connected \cite{easy},
so the only extra homotopy invariant is the boundary value itself.

The proof is an elementary application of obstruction theory. Since
$X$ is compact, it has a finite CW cellular structure
\cite{hat}. By subdividing cells
if necessary, we may assume that $\{p_1,p_2,\ldots, p_k\}\subset X^0$,
the $0$-skeleton of $X$. Let $K=X\times [0,1]$ and $L=X\times\{0\}\sqcup
X\times\{1\}\subset K$. Let $F^0:L\ra Y$ such that $F^0(\cdot,0)=f$ and
$F^0(\cdot,1)=g$. A (free) homotopy from $f$ to $g$ is then precisely
an extension of $F^0$ to some map $F:K\ra Y$. Since $f\sim g$, such an
extension $F$ exists by assumption. Let $\ol{K}^1=K^1\cup L$ and consider
the restriction of $F$ to $\ol{K}^1$, call it $F^1$. Clearly $F^1:\ol{K}^1\ra
Y$ has an extension to $K$, namely $F$ itself. Consider the canonical
CW structure on $\ol{K}^1$. It consists of 2 copies of $X$ connected by
$1$-cell spokes connecting $p\times 0$ with $p\times 1$ for each $0$-cell
$p\in X^0$. Since $f$ and $g$ are multiply based, the restriction of $F^1$ to
each spoke from $p_i\times 0$ to $p_i\times 1$, $i=1,2,\ldots,k$, defines
a closed loop in $Y$ based at $q_i$. If $Y$ is simply connected, each
such loop is nullhomotopic, so may be continuously shrunk to the constant
loop at $q_i$. Hence $F^1$ is homotopic to a map $\wt{F}^1:\ol{K}^1\ra Y$
which takes the constant value $q_i$ on each spoke $p_i\times 0$ to
$p_i\times 1$. But extendability of a map from a subcomplex of $K$ to the
whole complex $K$ depends only on its homotopy class \cite{ste}. Since
$F^1$ is extendable, so is $\wt{F}^1$, to $\wt{F}:K\ra Y$ say. $\wt{F}$
is precisely a homotopy between $f$ and $g$ through multiply based maps, 
which completes the proof.

The assumption that $Y$ is simply connected is necessary. In practice this
is almost always the case: $Y$ is usually a homogeneous space $Y=G/K$, which
is simply connected provided
$G$ is simply connected and $K$ is connected, by the homotopy exact 
sequence induced by the fibration
$K\hookrightarrow G\ra G/K$ \cite{exact}. Examples 
fitting this framework are the $O(3)$ and $\CP^N$ sigma models, the Skyrme
model and the Faddeev-Hopf model. Models of nematic liquid crystals are one
example where $Y$ is {\em not} simply connected. The order parameter 
represents a distribution of unoriented rods, so takes values in $\R P^2$.

Returning to the motivating example of the $O(3)$ sigma model on the
cylinder, here both $X$ and $Y$ are $S^2$ and there are two marked points
on each (not necessarily distinct on $Y$). Multiply based maps $f,g$ are
homotopic through maps preserving the boundary conditions if and only if
they are free homotopic, and hence if and only if they have the same
topological degree. This is a key underlying assumption of Rom\~ao's 
analysis of the system
\cite{rom}. In this case, the identification of $S^1\times\R$ with the twice
punctured $2$-sphere may be chosen to be holomorphic. 
If a path component of $(Y^X)_*$
contains a holomorphic map $X\ra Y$, then this map globally minimizes
 energy within its class, and hence is a stable static ``lump'' solution
by the usual Lichnerowicz-Belavin-Polyakov argument \cite{lic,belpol}.
Hence there is a moduli space of static $n$-lumps $\rat_{n,*}=
\rat_n\cap (Y^X)_*$ which allows Rom\~ao to study lump dynamics in some 
detail within the geodesic approximation of Manton. An interesting
question arises if $q_1=q_2$. Here $\rat_1\cap (Y^X)_*=\emptyset$:
there is no multiply based degree $1$ holomorphic map since such maps take
each value precisely once. Is there then no degree 1
energy minimizer? In fact there is not, though this clearly
 does not follow immediately. One direct way to see this is to construct
a divergent family of degree $1$ maps whose energy converges to the 
topological minimum value $\pi$. Use stereographic coordinates $z,W$ on 
$X,Y$, and let the marked points in $X$ be
$\pm 1$, and the coincident marked point in $Y$ be $0$. Then for each 
$\epsilon >0$, consider the map 
\beq
W_\epsilon(z)=\left\{\begin{array}{cl}
0& |z|>2\epsilon, \\
\epsilon^{2}z^{-1}& |z|<\epsilon, \\
h(\epsilon^{-1}|z|)\epsilon^{2}z^{-1}& \epsilon\leq |z|\leq 2\epsilon, 
\end{array}\right.
\eeq
where $h(\rho)$ is a 
smooth cut-off function which decays from $1$ to $0$ as $\rho$ varies from 
$1$ to $2$. Then $W_\epsilon$ has degree $1$, 
satisfies the basing conditions, and
 is holomorphic outside the small annulus, so by conformal invariance,
\bea
E[W_\epsilon]&<& \pi+\frac{1}{2}\int_{\epsilon\leq |z|\leq 2\epsilon}
 dz d\ol{z}\, \frac{|\cd W_\epsilon|^2+
|\ol{\cd}W_\epsilon|^2}{(1+|W_\epsilon|^2)^2}\nonumber \\
&<&\pi+\pi\epsilon^2\int_1^2 d\rho\, \rho\left[(\rho^{-1}h'(\rho)-
\rho^{-2}h(\rho))^2+\rho^{-4}h(\rho)^2\right]=\pi+c\epsilon^2
\eea
where $c>0$ 
is some constant. Hence $\inf E$ in the degree $1$ multiply based 
class is $\pi$, and this infimum cannot be attained.

\end{document}